\documentclass[5p]{elsarticle}
\usepackage{amsmath}
\usepackage{lineno,hyperref}
\modulolinenumbers[5]
\usepackage{color}
\usepackage{amssymb}
\usepackage[yyyymmdd,hhmmss]{datetime}
\def\be{\begin{eqnarray}}
\def\ee{\end{eqnarray}}

\def\({\left(}
\def\){\right)}
\journal{Carbon}
\def\({\left(}
\def\){\right)}

\def\be{\begin{eqnarray}}
\def\ee{\end{eqnarray}}

\begin{document}

\begin{frontmatter}

\title{The Largest Fullerene}

\author[a]{Michael Gatchell}
\author[a]{Henning Zettergren}
\author[b,c]{Klavs Hansen\texorpdfstring{\corref{cor1}}{}}
\ead{hansen@lzu.edu.cn,klavshansen@tju.edu.cn}
\cortext[cor1]{Corresponding author}
\address[a]{Department of Physics, Stockholm University, 106 91 Stockholm, Sweden}
\address[b]{Lanzhou Center for Theoretical Physics, Key Laboratory of Theoretical Physics of Gansu Province, Lanzhou University, Lanzhou, Gansu 730000, China}
\address[c]{Center for Joint Quantum Studies and Department of Physics, 
School of Science, Tianjin University, 
92 Weijin Road, Tianjin 300072, China}
\date{}

\begin{abstract}
Fullerenes are lowest energy structures for gas phase all-carbon particles for a range of sizes, but graphite remains the lowest energy allotrope of bulk carbon.
This implies that the lowest energy structure changes nature from fullerenes to graphite or graphene at some size and therefore, in turn, implies a limit on the size of free fullerenes as ground state structures. 
We calculate this largest stable single shell fullerene to be of size $N=1\times10^4$, using the AIREBO effective potential.
Above this size fullerene onions are more stable, with an energy per atom that approaches graphite structures.
Onions and graphite have very similar ground state energies, 
raising the intriguing possibility that fullerene onions could be the lowest free energy states of large carbon particles in some temperature range.
\end{abstract}


\end{frontmatter}

\section{Introduction}

The standard enthalpy of formation of the C$_{60}$ fullerene is 0.4 eV per atom \cite{KiyobayashiFST1993,SteeleJPC1992}.
The value is given relative to bulk graphite.
For isolated (gas phase) carbon clusters, in contrast, the fullerenes \cite{kroto1985}
are the lowest energy structures for small enough systems.
There must therefore be a limit to the size of gas phase fullerenes 
in equilibrium.

This work provides the results of a calculation of this size, involving a
comparison of the energies of the multilayer fullerene onions
\cite{UgarteNature1992} and the graphene \cite{NovoselovScience2004} 
or graphite structures.
In total, the icosahedral symmetry fullerenes, the compact graphite
structures, the fullerene onion structures, and graphene sheets 
were considered.
Clusters of fullerenes \cite{Hansen:2022}, which provide an
efficient way of producing large and mass selected all-carbon gas
phase systems, can be ruled out as candidates for lowest energy 
structures by the known values of 
fullerite binding energies.
They will therefore not be considered here further.

\section{Computational method}

\begin{figure*}[htb]
\centering
\includegraphics[width=0.7\textwidth]{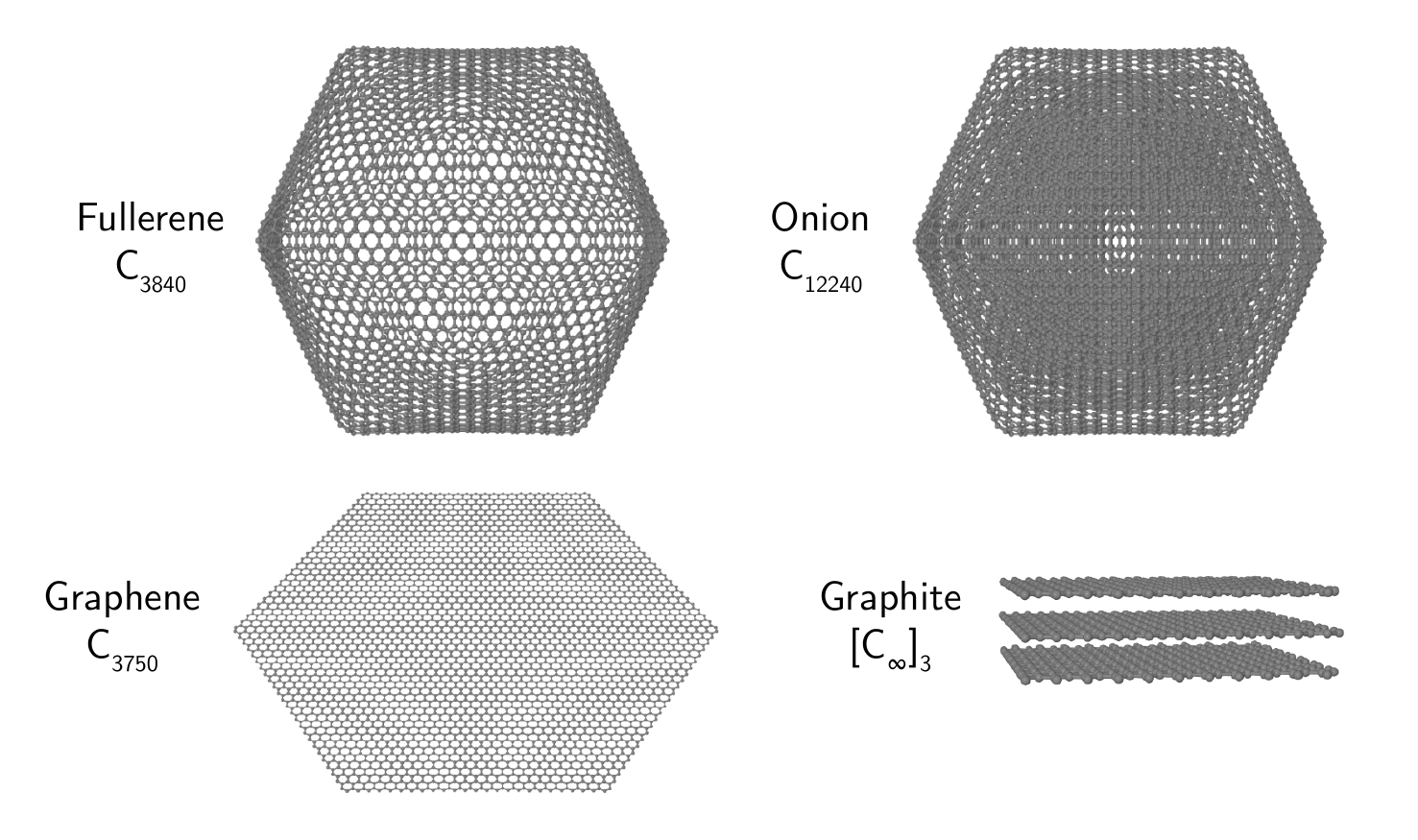}
\caption{\label{structures}
The ground state configurations of structures considered in this work. 
From left to right, top row: an icosahedral fullerene structure
C$_{3840}$), a nested fullerene structure, denoted onions (C$_{12240}$,
consisting of 8 layers); bottom row: a finite monolayer of graphene
(C$_{3750}$), and a truncated portion of an infinite three-layer graphite
particle.}
\end{figure*}

Models based on structural motifs have been shown to 
accurately predict the relative stability of fullerene isomers ranging from C$_{60}$ to icosahedral fullerenes containing up to 6000 carbon atoms \cite{Austin1995,ItohPRB1996,Achiba1998, Cioslowski2000, Alcami2007, WangPCCP2017}.
In these works, the results of Density Functional Theory calculations are typically used as input to arrive at simple expressions for the relative fullerene energies. 

The same strategy is followed in this work.
The system sizes that are relevant in this study are orders of magnitude larger and will render any attempt of a full quantum mechanical calculation unfeasible.
Instead, the empirical AIREBO potential by Stuart \emph{et al.}\ was used to describe the interactions, both of intra- and intermolecular nature \cite{Stuart:2000aa}.
This is a reactive manybody potential where the interaction strengths and bond angles depend on the local environments of the participating atoms. Long-range dispersion forces are also included in the definition of the AIREBO potential. 
It has been used previously to accurately model graphene and fullerenes in numerous studies \cite{Irle:2006te,polym6092404,Seo:2020vq} and generally compares favorably with more advanced theory and experimental data \cite{polym6092404}.

The initial structures of graphene layers and fullerene cages were generated 
using a custom code that built up graphene sheets with a D$_{6\text{h}}$ symmetry
(see Figure \ref{structures}). From these, the 20 triangular faces of icosahedral
(I$_{\text{h}}$ symmetry) fullerenes could be formed from Coxeter constructions
with $(m,n)=(1,1),(2,2),(3,3)\dots$
\cite{Coxeter:1971ue,Fowler:2007vr,Siber:2020wa, Schwerdtfeger2015}. 
Such icosahedral fullerenes consist of $60n^2$ atoms, where 
$n$ is the same positive integer as used in the Coxeter 
constructions.

The fullerene onions that have been observed experimentally
are more spherical than the lowest energy single shell 
fullerenes (see \cite{Schwerdtfeger2015} and references 
therein).
Calculations \cite{Bates1998} suggest that this is due to 
C$_2$-emission that introduce one heptagon and an additional pentagon.
However, as the binding energy per atom for these two 
classes of onions is very similar for species containing 
more than thousand atoms (see Fig.\ 6 in \cite{Bates1998}), 
and as the introduction of a heptagon combination in any
case increases 
the energy, it is justified to consider the icosahedral 
structures here.

Graphite structures were formed by stacking graphene 
layers with a small 
offset in the $xy$-plane (equal to the length of a 
C-C bond) between each
layer so that each hexagonal ring had a carbon atom 
from the two surrounding
layers placed directly above its center and such that 
every second layer
shared the same $xy$-coordinates. 
Infinite layers were modeled using periodic boundary 
conditions in the $xy$-plane and the asymptotic graphite 
limit with an infinite number of
layers was reached by also including periodic boundaries 
along the $z$-axis.

The multilayered onion structures were built
by placing successively larger fullerene
structures around all of the smaller
icosahedral fullerenes, starting
with C$_{60}$ as the innermost shell.
The sizes of such onions is given by the
sum $N=60 \times \sum_{n=1}^{n_{\rm o}} n^2$,
where $n_{\rm o}$ is the Coxeter index of the
outermost fullerene layer.
The fullerene layers were structured with
their symmetry axes aligned. 
The input geometries were optimized under
the AIREBO force field using the LAMMPS
molecular dynamics software package
\cite{Plimpton:1995aa} to provide the
ground state structures and energies. 
Figure \ref{structures} shows select
optimized structures of the geometries 
considered here.
A complete archive of the optimized structures is available as
well \cite{largest_fullerene_data}.

\section{Results}

\begin{figure*}[htb]
\centering
\includegraphics[width=0.9\textwidth]{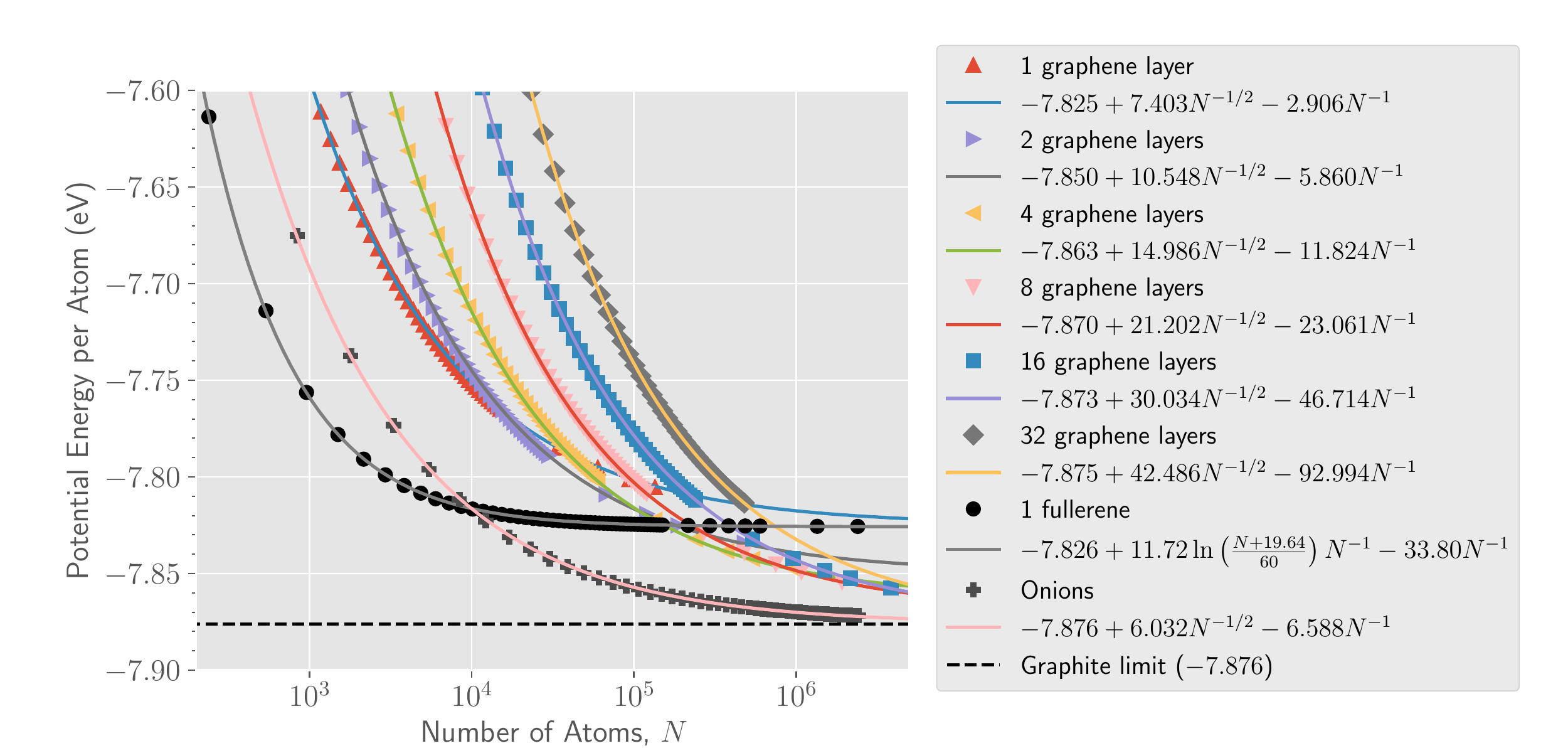}
\caption{\label{Fig1LargestFullerene}
The ground state energy per atom for the different structures calculated here in eV. Fits to each class of of particles are shown as solid lines with the parameters given in the legend. The limit energy of bulk graphite consisting of a infinite number of infinitely large graphene layers is presented as the dashed line.
}
\end{figure*}

The ground state potential energy per atom for the different 
structures are shown as a function of size in Figure 
\ref{Fig1LargestFullerene}. 
The smallest system considered here is a single C$_6$ ring (not 
shown in the figure) while the largest structures that we have
explicitly studied contain millions of atoms. 
As seen from the figure, the single shell fullerene allotrope 
is the lowest energy structure type for systems containing 
up to approximately $10^4$ atoms. 
Above this size the energy of the multi-layer onion fullerene 
structures dives below that of the single shell fullerenes 
which has already reached energies close to the asymptotic 
value of $-7.83$\,eV per atom. 
The smallest onion that we identify in our calculations with 
a lower potential energy per atom than a comparably sized 
single shell fullerene is the C$_{12240}$ cluster that consists 
of 8 fullerene layers (shown in figure \ref{structures}).

Figure \ref{Fig1LargestFullerene} also gives least square fits of energies vs. size to facilitate  comparisons between the different species. 
For graphene and the graphitic and onion-like structures, the functional form of the fit is given by 
\be
E(N,n)/N = E_1(n) + E_2(n)/\sqrt{N} + E_3(n)/N,
\label{eq:fit_sqrt}
\ee
where $N$ is the total number of atoms in the system, $n$ is the number of graphene layers (not relevant for the onions), while $E_1$, $E_2$, and $E_3$ are free parameters that depend on $n$.
The trends are well-described by this functional form. 
The $E_2$ and $E_3$ terms summarize the contributions from surface energies, e.g., the edge effects in the finite graphene layers, which become negligible in the continuum limit.
For the icosahedral fullerenes we find that this form gives a poor representation of the scaling of the potential energy with size.
These compounds do not have edges and instead the next-to-leading order term (in the total energy) was taken to be of logarithmic form. 
This has been used previously and found to give a good fit for the strain that arises from the curvature introduced by the 12 pentagons \cite{Tersoff:1992wy,SiberEPJD2006,WangPCCP2017}. 
The relatively slowly varying logarithmic form should be compared with the value for a perfect sphere which in the continuum limit will have a size-independent total strain energy \cite{GuanPRB2014}.
The onion energies include the higher strain from the smaller Russian doll fullerenes.
Including this and the effective surface term from the graphite-like interaction of the layers adds up to the term proportional to the square root of the total number of atoms, at least to leading order.

The dashed line in Figure \ref{Fig1LargestFullerene} shows the potential energy limit of ideal bulk graphite within the AIREBO model. 
The calculation using periodic boundary conditions (PBCs) to simulate graphite consisting of an infinite number of infinitely large graphene layers represents the bulk ground state of carbon with no external pressure applied. 
The limit for finite numbers of graphene layers were also determined for systems consisting of up to 256 layers using PBCs in only two dimensions. 

Of the systems shown in Figure \ref{Fig1LargestFullerene}, only the onion structure asymptotically approaches the graphite limit within the uncertainties of the fitted curves. 
The asymptotic limits for the different graphene, fullerene, and onion particles, determined from the fits shown in Figure \ref{Fig1LargestFullerene} and the calculations using PBCs, are given in Table \ref{tab:limits}.

\begin{table}[htb]
\caption{Asymptotic limits on the potential energy per atom ($E_1$) determined from the fits in Figure \ref{Fig1LargestFullerene} (center column) and from calculations using periodic boundary conditions to achieve infinitely large layers (right column). 
The fitted values are given with their respective uncertainties.
Here and in the following the error bars are calculated as statistical.}
\vspace{0.3cm}
  \label{tab:limits}
  \begin{tabular}{l|cc}
Structure & Fit Limit (eV)  & PBC Limit (eV) \\ 
    \hline
1 Graphene Layer & $-7.82497(8)$ & $-7.82569$ \\ 
2 Layers & $-7.84985(9)$ & $-7.84931$ \\ 
4 Layers & $-7.86322(7)$ & $-7.86275$ \\ 
8 Layers & $-7.86970(7)$ & $-7.86948$\\ 
16 Layers & $-7.87307(6)$ & $-7.87285$ \\ 
32 Layers & $-7.87472(6)$ & $-7.87454$ \\ 
64 Layers & - & $-7.87538$ \\
128 Layers & - & $-7.87580$ \\
256 Layers & - & $-7.87601$ \\
Fullerene & $-7.82577(2)$ & - \\ 
Onion & $-7.87611(6)$ & -  \\ 
Bulk Graphite & - & $-7.87622$
\end{tabular}
\end{table}

\begin{figure}[htb]
\centering
\includegraphics[width=\columnwidth]{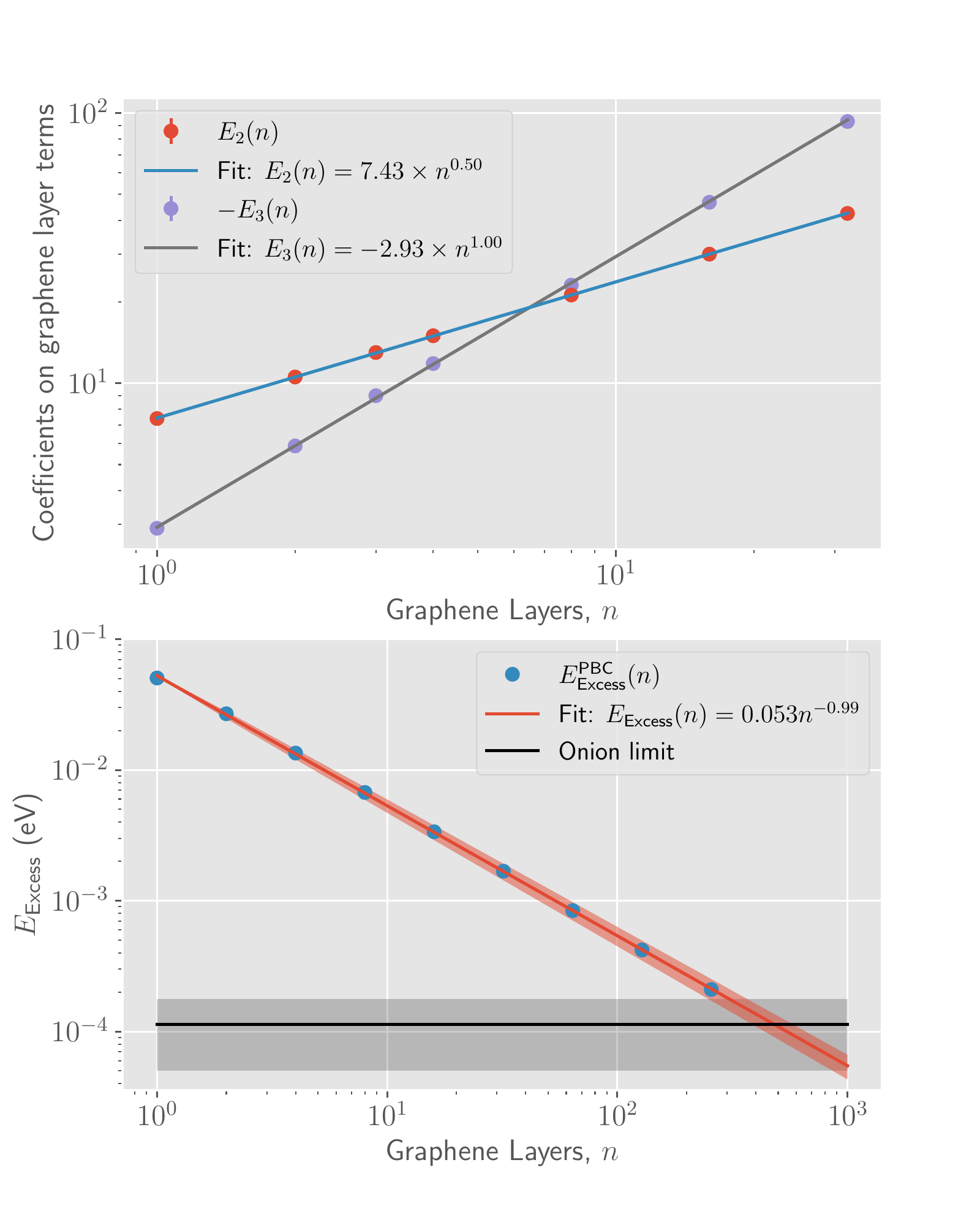}
\caption{\label{GrapheneLayersFits}
Top panel: The coefficients to the constant and $N^{1/2}$ terms from the fits to eq.\ (\ref{eq:fit_sqrt}), $E_1$ and $E_3$ respectively, for different numbers of stacked graphene layers. Bottom panel: The scaling of the asymptotic potential energy, $E_2$, of stacked graphene layers as a function of $n$ together with a fitted power law expression and the corresponding limit for onion structures. 
The shaded areas in both panels show the $1\sigma$ statistical uncertainties of each curve.
}
\end{figure}

The parameters determined from the fits, $E_1$, $E_2$, and $E_3$, can be used to extract a scaling relation for the potential energy of graphitic particles of arbitrary sizes. Figure \ref{GrapheneLayersFits} shows fits made to these parameters as functions of the number of layers, $n$. 
The fits made in the upper panel, where the exponents are free parameters, suggest $E_3$ scales linearly with $n$ while $E_2$ scales as the square root of $n$ over the range of particle sizes modeled. 
In the lower panel of Figure \ref{GrapheneLayersFits} we show the scaling of the asymptotic potential energy limit ($E_1$) in terms of the so-called excess energy. 
The excess energy, $E_{\text{Excess}}$, is defined as the difference per atom in potential energy between a given particle and that of bulk graphite. Here, the limits obtained for $n$ graphene layers from the PBC calculations, $E_{\text{Excess}}^{\text{PBC}}(n)$, are shown as the different data points.
In the figure, the asymptotic limit for the onion structures is shown as a constant while the corresponding excess energy for up to 256 stacked graphene layers is shown as the blue points. 
The red curve shows a fit of a powerlaw to these point that allows the $E_1(n)$ parameter to be scaled to sizes larger than those explicitly modeled. The exponent from this fit, $p=-0.994\pm0.004$, is consistent with a $1/n$ scaling of $E_{\text{Excess}}$. The same fit using the ``fit limit'' values from Table \ref{tab:limits} result in an exponent of $p=-1.019\pm0.009$.
Noteworthy is that the fitted curve crosses the curve for the onions structures at $n=480$.
This is an estimate of the fewest number of graphene layers required to give a structure with a lower potential energy than the onions, made with the fit parameters explained below..

\begin{figure}[htb]
\centering
\includegraphics[width=\columnwidth]{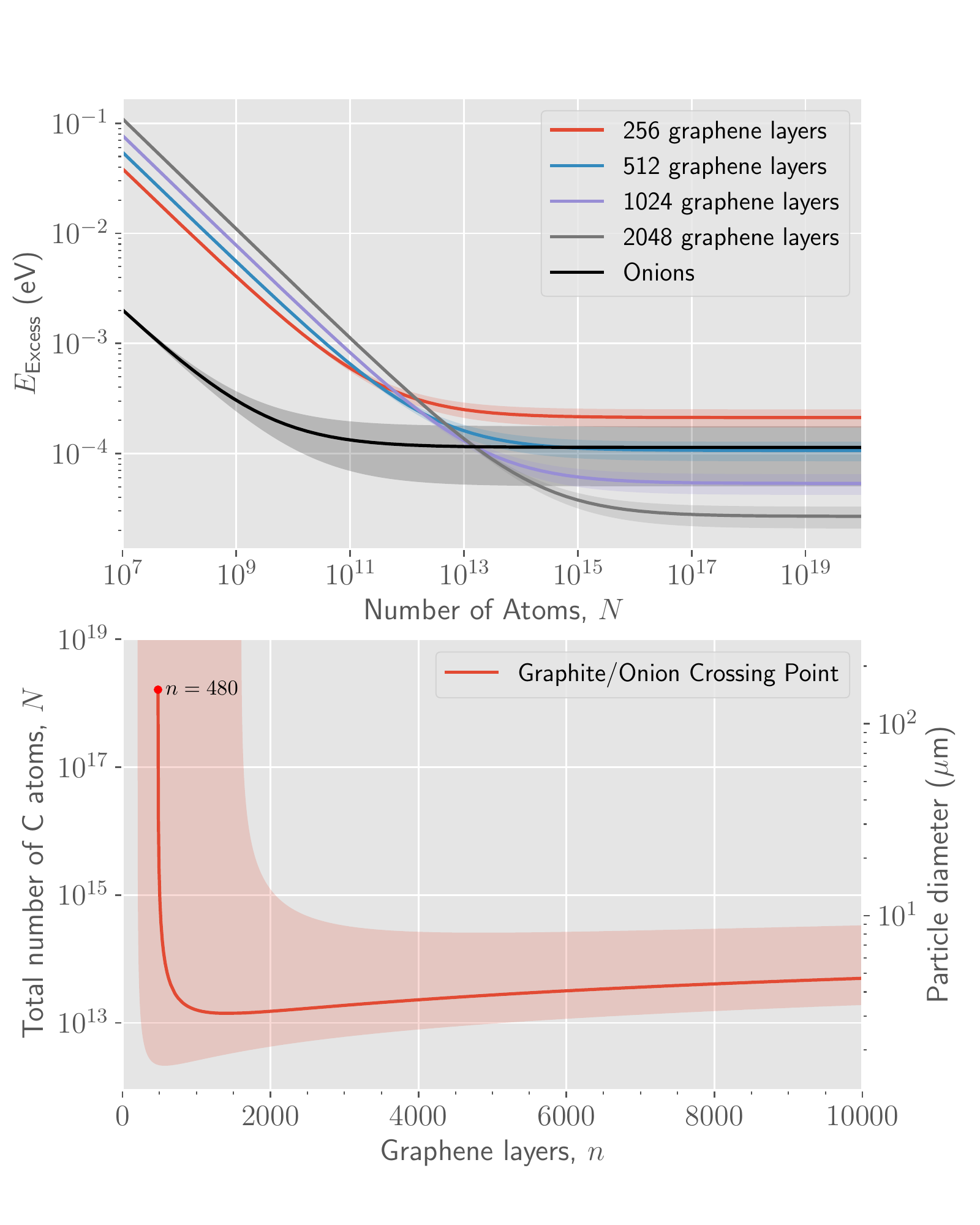}
\caption{\label{GraphiteSizes} Top pane: Extrapolated potential energy curves for fullerene onions and graphitic particles consisting of different numbers of layers. The first graphitic curve to cross below the onion curve corresponds to $n=480$ graphene layers (not shown) at $N\approx10^{18}$ atoms in total. Bottom panel: The particle sizes for which the potential energy of a graphitic particle is lower than that of an onion particle with the same size $N$ as function of the number of graphene layers, $n$. The shaded areas in both panels show the $1\sigma$ statistical uncertainties of each curve.
}
\end{figure}

From the fitted parameters shown in Figure \ref{GrapheneLayersFits} we 
can derive a single expression for how the potential energy per atom 
of a graphite particle scales with the total number of atoms ($N$) 
and number of graphene layers ($n$). 
The expression we arrive at is
\begin{align}
\label{graphiteparticle}
    &E_\text{graphite}(n,N)/N =  \nonumber \\
    &= -7.87622 + 0.053n^{-0.99} +7.43\sqrt{n/N} -2.93n/N,
\end{align}
where $E_\text{graphite}$ is given in units of eV. 
Optimizing the energy with respect to the layer number, $n$, gives 
the ground state minimum energy.
An approximate value is found for large $N$ by solving for the 
minimum with only the second and third terms 
\be
n \approx 0.0588..N^{1/3}.
\ee
This is a rather small number of layers, reflecting the large 
cost of creating an edge on a graphene sheet compared to separating 
graphene sheets.
The parametrization in Eq.(\ref{graphiteparticle}) allows us to
extrapolate the ground state potential energies for 
particles consisting of stacked graphene sheets to sizes larger that 
we are able to explicitly simulate without using PBCs. 
An example of this, with the minima calculated numerically with the 
full expression in Eq.(\ref{graphiteparticle}), is shown in the upper
panel of Figure \ref{GraphiteSizes}. 
There, the excess energy per atom is plotted for $n=256$, 512, 1024,
and 2048 layers of graphene, as well as the corresponding curve for
fullerene onion structures. 
Due to the  uncertainties in these calculations, indicated by 
the shaded areas in the figure, the minimal number of 480 graphene 
layers given above could be as small as approximately 200. 
The curve for 480 graphene layers (not shown) crosses the onion 
curve at a particle size of $N=10^{18}$ atoms (only a single 
significant digit is given in consideration of the calculation 
uncertainties). 
However, for particles consisting of more layers, the crossing point occurs at smaller total sizes, as seen in the curves for 512, 1024, and 2048 layers. In the lower panel of Figure \ref{GraphiteSizes} we show the crossing point as a function of the number of graphene layers in a graphite particle. 
Starting from 480 layers, the corresponding crossing point is seen 
to take place at smaller particle sizes until a minimum is reached 
at 1160 layers, where the cross over points between onions and graphite 
takes place for $10^{13}$ atoms. 
This is then the estimate of the size of the largest stable onion. 
In terms of particles size, this corresponds to a spherical particle 
with a diameter of a few micrometers.
After this there is a slight increase in the cross-over particle size with 
increasing number of graphene layers. 

%

\section{Summary and Discussion}

The results here show that the minimum energy
structure of free carbon particles is, in order of 
increasing size, single shell fullerenes, onions, and 
graphite. 
The coefficients to the leading order terms of graphite and
onion structures differ for two reasons.
One is the inter-layer interaction.
This is a free parameter for the graphite sheets and can be 
optimized unconstrained, whereas the onion structures are 
limited to the radii defined by the closed shells number of
atoms equal to $60 n^2$. 
The energy penalty of this distance effect is minor, however,
seen also from the fact that the optimal distance between 
graphite layers is close to the
distance given by the radii of two consecutive shells in the
onion. 
The close match makes the magnitudes of the leading order 
terms of the onions and the graphite very similar.
Another diffeerence is the vibrational spectra.
These will add different vibrational zero point energies to 
the calculated structures.
It has not been possible to treat this question here due to 
the very large particle numbers involved. 

The limitations of the numerical calculations notwithstanding, 
we can nevertheless draw some conclusions and make some 
conjectures.
One aspect concerns the expected reduction in the HOMO-LUMO 
gap with increasing size predicted by No\"{e}l \emph{et al.}\
\cite{NoelPCCP2014}.
The authors extrapolated their calculated gap beyond their 
highest calculated size of $6\times 10^3$ to find that 
the gap vanishes above $N=7\times 10^4$.
This is above the calculated largest fullerene size and 
raises the question of the use of these fullerenes for 
electronics in analogy to the uses envisioned for graphene.

The implications of these findings could be important for 
a better understanding of the carbon inventory of interstellar 
and circumstellar environments. The crossover point from onions to graphite in our model occurs at particle sizes on the order of micrometers. This is of the same order of magnitude as the upper limit of interstellar dust grains \cite{Weingartner:2001aa}, which could suggest that onion-like structures are prevalent in these regions. Previous experimental work has indeed suggested that fullerene onion particles could explain certain interstellar extinction features \cite{Tomita:2004aa}, lending support to this possibility.

The main unresolved question of this work is how the energies calculated compared to the reaction barriers between the different allotropes.
This will remain a topic for future study. 
We should also mention the potential role of imperfections in the structures, as well as the possibility of other, more exotic carbon structures \cite{Meirzadeh:2023aa}.
These open questions make it difficult to formulate any strong experimental predictions. 
We will express a guarded optimism concerning conversion rates, however. 
In contrast to expectations, carbon dimers have been shown to attach barrier-less to C$_{60}$ in gas phase, and to be incorporated into the cage \cite{DunkNatComm2012}.
Hence the combined process of bond breaking and bond formation in a conversion may only experience barriers for the bond breaking part.

\section{Acknowledgements}

KH acknowledges support form the National Science Foundation of China with 
the grant 'NSFC No. 12047501' and the 111 Project under Grant No. B20063 
from the Ministry of Science and Technology of People's Republic of China.

HZ and MG acknowledge financial support from the Swedish Research Council 
(contracts 2020-03437 and 2020-03104) and from the project grant "Probing charge- and mass- transfer reactions on the atomic level"  (2018.0028) from the Knut and Alice Wallenberg Foundation.

This publication is based upon work from COST Action CA18212 - Molecular Dynamics in the GAS phase (MD-GAS), supported by COST (European Cooperation in Science and Technology)


\end{document}